\documentclass{aa}  

\usepackage{graphicx}
\usepackage{txfonts}
\usepackage{soul}
\usepackage[normalem]{ulem}
\RequirePackage{xcolor}
\usepackage[
  breaklinks = true,
   colorlinks = true,
  urlcolor   = blue,
  citecolor  = blue,
  linkcolor  = red, ]{hyperref}
\definecolor{teal}{rgb}{0.0, 0.5, 0.5}
\definecolor{lava}{rgb}{0.81, 0.06, 0.13}
\graphicspath{ {./figures/} }

\def\kms{\hbox{km$\;$s$^{-1}$}}

\usepackage{multimedia}
\begin{document}

%\title{High-resolution observations of recurrent jets changing direction due to a moving reconnection site}
\title{High-resolution observations of recurrent jets from an arch filament system}

   \author{Reetika Joshi
          \inst{1,2}
          \and
          Luc Rouppe van der Voort
          \inst{1,2}
          \and
          Brigitte Schmieder\inst{3,4,5}
          \and
          Fernando Moreno-Insertis\inst{6,7}
          \and
          Avijeet Prasad\inst{1,2}
          \and
          Guillaume Aulanier\inst{8,1}
          \and
           Daniel Nóbrega-Siverio\inst{6,7,1,2}
          }

   \institute{Rosseland Centre for Solar Physics, University of Oslo, P.O. Box 1029 Blindern, N-0315 Oslo, Norway\\
   \email{reetika.joshi@astro.uio.no}
    \and
    Institute of Theoretical Astrophysics, University of Oslo, P.O. Box 1029 Blindern, N-0315 Oslo, Norway    
         \and
    LIRA, Observatoire de Paris, Universit\'e PSL, CNRS, Sorbonne
    Universit\'e, Universit\'e de  Paris,
    %Diderot, Sorbonne Paris Cité, 
    5 place Jules
    Janssen, F-92195 Meudon, France
    \and
    Centre for mathematical Plasma Astrophysics, Dept. of Mathematics, KU Leuven, 3001 Leuven, Belgium
     \and
    School of Physics and Astronomy, University of Glasgow, Glasgow G12 8QQ, UK
    \and
    Instituto de Astrof\'{\i}sica de Canarias, E-38200 La Laguna, Tenerife, Spain 
    \and
    Departamento de Astrof\'{\i}sica, Universidad de La Laguna, E-38206 La Laguna, Tenerife, Spain 
    \and
    Sorbonne Universit\'e, École Polytechnique, Institut Polytechnique de Paris, Observatoire de Paris - PSL, CNRS, Laboratoire de physique des plasmas (LPP), 4 place Jussieu, F-75005 Paris, France
             }

   %\date{Received September 15, 1996; accepted March 16, 1997}

% \abstract{}{}{}{}{} 
% 5 {} token are mandatory
 
  \abstract
   {Solar jets are collimated plasma ejections along  magnetic field lines 
observed in hot (EUV jets) and cool (chromospheric surges) temperature diagnostics. Their trigger mechanisms and the relationship between hot and cool jets are still not completely understood.}
   {We aim to investigate the generation of a sequence of active region solar jets and their evolution from the photospheric to the coronal heights using multi-thermal observations from ground-based and space-borne instruments.}
{Using the synergy of high spatial and temporal resolution observations by the Swedish 1-m Solar Telescope (SST), along with the Solar Dynamics Observatory (SDO), we analyze a sequence of solar jets originating  
in  a mixed polarity region between the leading and following sunspots of an active region.
Kinematics of these jets is investigated using the spectra from the SST observations.
 We use a Non-Force-Free Field (NFFF) extrapolation technique for deriving the magnetic field topology of the active region.}
   {A mixed  polarity region is formed over a long period (24 hours) with persistent magnetic flux emergence. This region has been observed as an arch filament system (AFS) in chromospheric SST observations.
   In this region, negative polarities surrounded by positive polarities create a fan-surface with a null point at a height of 6~Mm detected in the NFFF extrapolation. 
   SST observations in H$\beta$ spectral line reveal a large flux rope over the 
   AFS and moving from the North to South, causing successive EUV and cool jets
   to move in the East-West direction and later towards the South along the long open loops.
   }
   {The high resolution SST observations (0\farcs038 per pixel) resolve the dark area observed at the jet base and reveal the existence of an AFS with an extended cool jet  which may be the result of a peeling-like mechanism of the AFS.  
   Based on the combined analysis of SST and AIA observations along with extrapolated magnetic topology, it is suggested that the magnetic reconnection site may move southward by approximately 20~Mm until it reaches a region where the open magnetic field lines are oriented North-South.}


   \maketitle
%
%------------------------------------------------------------------
%Figure 1

\section{Introduction}
High resolution observations from ground and space have revealed that the Sun is continuously generating ejections starting in the low atmosphere and reaching the corona. The generic solar jets are observed as multi-thermal collimated plasma flows along the magnetic field lines \citep[see e.g., ][]{Shibata1992, Canfield1996, Nistico2009, Chandra2015, Joshi2017}.  Solar jets are observed in a wide range of wavelengths in the solar spectrum and have a great variety of sizes, from the very small scale, similar to spicules \citep{Tian2014}, to large scale eruptions leading to coronal mass ejections \citep{Wang2002,Shen2012,LiuJiajia2015,Joshi2020ApJ}.
They are observed in different solar domains like, for instance, in coronal holes \citep{Panesar2018}, active regions, penumbra \citep{Liu2022}, light bridges \citep{Asai2001,Robustini2016}, and  even in the quiet sun  or at the pole  
\citep{Nistico2009,Raouafi2014}.  Up- and downflows  are commonly observed in solar jets \citep{Schmieder1997, Cheung2015, Joshi2024}. Several reviews summarize the observations and the simulations from the past to the present \citep{Raouafi2016, Schmieder2022}. 

 Solar coronal jets have been classified as standard and blowout jets \citep{Moore2010}. The triggering mechanism behind both of these classes of jets is  magnetic reconnection which allows to release the stored magnetic energy, converting it into kinetic and thermal energy flux in the reconnection outflows. Magnetic reconnection in tiny regions showing small-scale jetting activity (called jetlets) is also proposed to be a source for heating the solar corona and to act as a driver of the solar wind \citep{Raouafi2014, Panesar2018}. There are different theoretical explanations about the drivers for the jets, which may differ depending on the presence of emerging magnetic field, magnetic flux cancellation or eruption of a mini-filament.
 %{\bf However, \citet{Muglach2021} presented a statistical study of solar jet observations and claimed that none of the existing jet simulations are able to explain the observational signatures of converging flows and magnetic flux cancellation.}

The theoretical idea of the standard jet eruption model was first proposed by \citet{Heyvaerts1977} and further developed by \citet{Shibata1992}, and  later numerically simulated by several authors in 2D \citep{Yokoyama1995, Yokoyama1996} and 3D \citep{Moreno2008, Moreno2013}.
In these models, closed magnetic arch-like  bipolar flux  emerges into the ambient unipolar field and then reconnection occurs between the unipolar field and the branch of the arch with the polarity opposite to the unipolar field.  
This reconnection outflow leads to a fast-mode shock with the plasma flow upward along the open magnetic field lines and creates a jet.
Observations of anemone jets observed with Hinode \citep{Tsuneta2008,Suematsu2008}  belong to this class of standard jets, as explained by \citep{Nishizuka2008, Nishizuka2011}.
In chromospheric H$\alpha$ observations these flux emerging regions can be observed as arch filament systems \citep[AFS;][]{Bruzek1967}. AFS are defined as a system of very low and dark filaments 
\citep{Mandrini2002, Ma2015}. These systems have a lifetime of several days; however, individual arches show a lifetime of a few tens of minutes. It has been observed that the hot flare loops overlie these AFS at coronal heights \citep{Alissandrakis1990, Malherbe1998}.

X-ray jets are frequently accompanied by cool plasma jets 
called solar surges; those emerge as roughly straight threads of chromospheric dark material \citep{Canfield1996, Schmieder1996, Wang1998SP, Chae1999, Mandrini2002, Uddin2012}. 
Surges can be detected in hot coronal channels, such as:  171 \AA, 193 \AA, 211 \AA\ due to the fact that they block the EUV emission coming from layers below them \citep{Anzer2005,Heinzel2008}. Therefore  surges appear as dark areas 
alongside the hot jets in these AIA EUV channels \citep{Nelson2019, Joshi2020}. 

The other class of jets called   
blowout jets are identified by their width which is as large in the ejection part as at the base \citep{Moore2010,Moore2013}. 
Blowout jets have been observed with two phases; the first phase is similar to the standard model, but it is followed by a second phase characterised by the presence of an elongated darkening in the EUV images, interpreted as a mini-filament \citep{Sterling2015,Sterling2022,Panesar2016,Panesar2022, Kumar2019}. 
These rising filaments are assumed to bring free  energy due to shear or twist and reconnect with the ambient magnetic field, thereby initiating blowout jets \citep{Wyper2019,Sterling2022,Kumar2023}. Many jets have helical motions  and their twist can be transported from a nearby flux rope to the jet 
\citep{Joshi2020FR, Joshi2021}. 
The role of  
twist for unfurling the blowout jet material has also been obtained in several numerical experiments \citep{Pariat2009, Pariat2010, Moreno2013, Archontis2013, Fang2014, Pariat2015, Wyper2017}.

Using multi instrument observations, 
several studies on jets have been done to explain the  twist in mini filaments at the base of blowout jets \citep{Cheung2015,Tiwari2018}, bilateral flows at the reconnection points \citep{Ruan2019}, reconnection in null points or in bald patch regions \citep{Joshi2020FR}, and the formation of plasmoids (blob-like plasma ejections or bright kernels) \citep{Singh2012,Zhang2014,Rouppe2017,Bogdanova2018,Chen2022}.
These bright kernels trace the bilateral flows forming at the junction  of the arch and the spire of the jet and commonly have a higher density and temperature than the jet. Numerical experiments of the jet process propose the tearing mode instability to form the plasma islands associated with the bright kernels \citep{Yang2013, Ni2016, Xiaohong2023}. 

In the present study, using high spatial and temporal resolution observations with the Swedish 1-m Solar Telescope \citep[SST;][]{Scharmer2003} together with observations from the SDO satellite, we analyze a sequence of jets originating from an active region in which emerging magnetic flux (EMF) is continually appearing. The moving magnetic patches associated with the EMF episodes form a  favorable site for jet launching in the close vicinity of the leading sunspot. Small B-class flares, with one of them reaching the 
B8.8-class  as  recorded by the GOES satellite,  
were related to recurrent  ejections of plasma blobs and jets from these emerging regions. 
We present, in Sect. \ref{sec:obs}: the overview of the jet observations; in Sect. \ref{sec:recurrent}: the recurrent jet environment and the jet dynamics; in Sect. \ref{sec:NFFF}: the non force-free field extrapolation results. 
In Sect. \ref{sec:discuss} we discuss the trigger and occurrence of the jets according to the magnetic configuration and the drifting of the null point location. 
\begin{figure*}[t!]
   \centering
      \includegraphics[width=\textwidth]{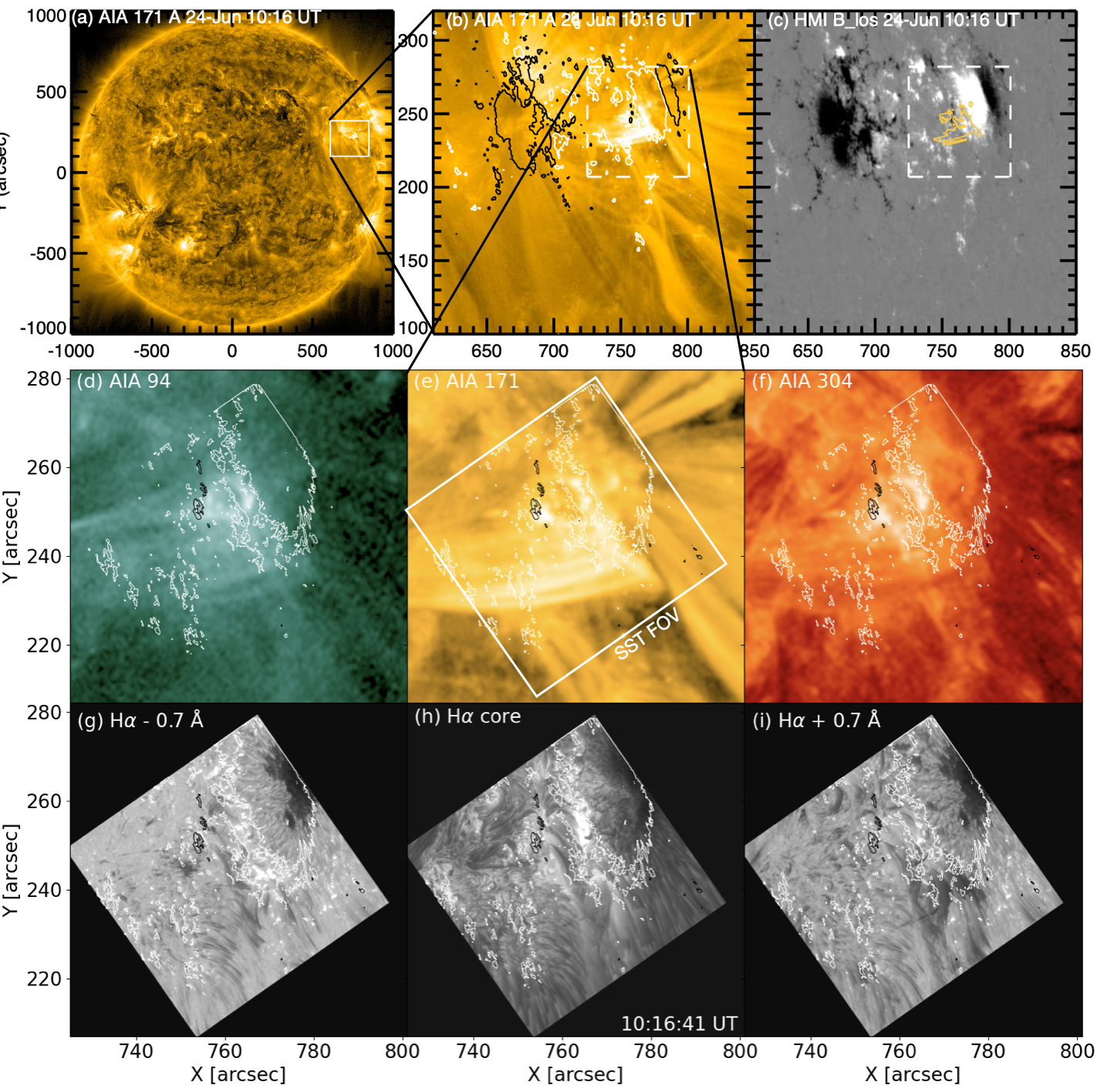}
   \caption{General overview of AR 13038 where the location of the AR is marked with a white rectangular box in the full disk AIA~171~\AA\ image (panel a) and the focused FOV is shown in panels b and c. The white box in panel (b-c) is the FOV for the middle and last row, showing the maximum FOV which was covered during the coordinated observations by SST.
   The straight enhancement of brightening in 171~\AA\ (yellow contours of AIA 171 in panel c) 
   is along with the polarity inversion line between the main positive polarity and the emerged negative polarity (N1, N2, N3 in Fig.~\ref{FigHMI_SST}). 
   The white and black contours are labelling the magnetic field strength of $\pm$500~G.
    The middle row presents the AIA observations aligned to SST observations at a specific time and the AIA spatial resolution is blown up to the SST spatial resolution. An animation is attached online (see \url{http://tsih3.uio.no/lapalma/subl/afs/fig1_AIA_SST.mp4}) with this figure to show the evolution of the jet region from 09:47 to 11:30 UT. 
   }
\label{AIA_HMI}
    \end{figure*} 
 
    \begin{figure*}[t!]
   \centering
   \includegraphics[width=\textwidth]{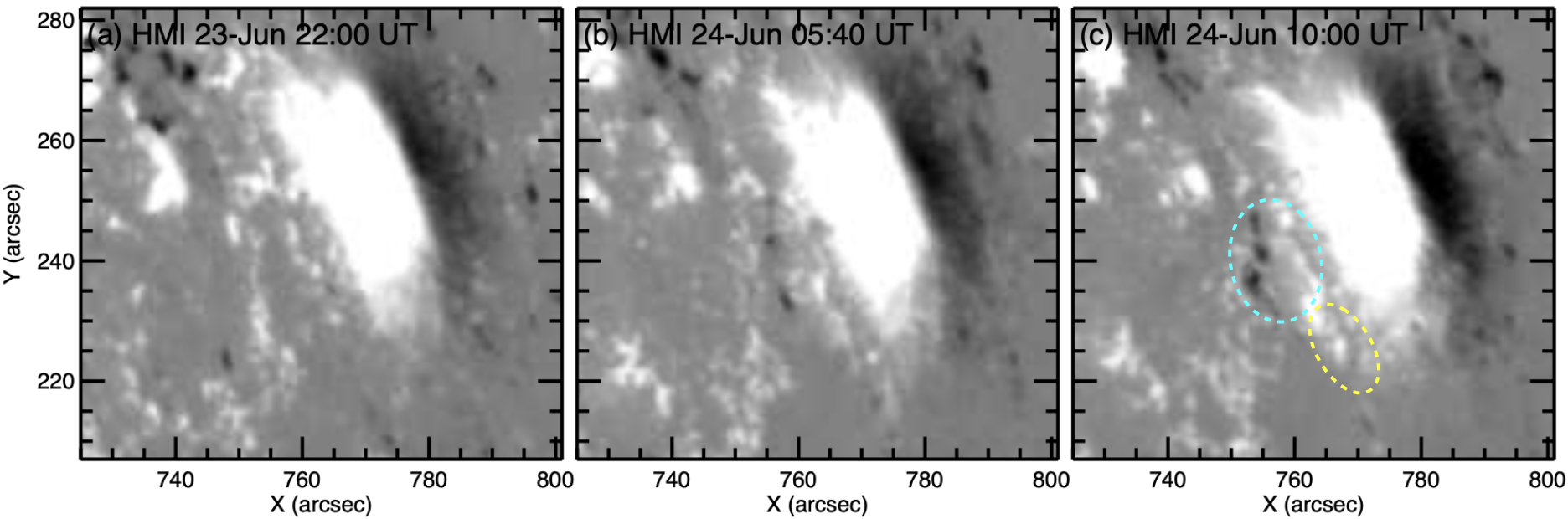}
   \caption{Evolution of the emerging flux on the East side of the leading positive sunspot over 12~h duration. This sunspot acts as a source of flux emergence and emerged polarity patches move to form the regions marked by the two ovals in panel (c). 
   The negative polarity on the West side of the sunspot is an artifact due to the projection effect. The threshold of the magnetic field is $\pm$ 500 G.
   }
              \label{FigHMI_SST}%
    \end{figure*}

\section{Observations}
\label{sec:obs}
\subsection{Ground-based observations}
Observations of the AR NOAA 13038 (N16 W61) were taken during the coordinated observational campaign involving the  
SST and SDO in June 2022. The observations were recorded using two instruments at the SST: the CRisp Imaging SpectroPolarimeter \citep[CRISP;][]{Scharmer2008}, and the 
CHROMospheric Imaging Spectrometer \citep[CHROMIS;][]{Scharmer2017}. CRISP and CHROMIS are Fabry--P\'erot tunable filter instruments equipped with dual etalon systems \citep{Scharmer2006}. The CRISP instrument has the capability to measure the linear and circular polarisation in photospheric and chromospheric spectral lines. 
With CRISP, the observations were taken in H$\alpha$ (31 line positions from $-$1500 to +1500~m\AA), \ion{Ca}{ii} 8542~\AA\ (20 line positions from $-$1680 to +2380~m\AA), and \ion{Fe}{i} 6173~\AA\ (14 line positions from $-$320 to +680~m\AA) at a cadence of 40~s. The \ion{Ca}{ii} and \ion{Fe}{i} observations were recorded in spectropolarimetric mode. The CHROMIS data were taken in H$\beta$ 4861~\AA\ (27 line positions between $\pm$2100~m\AA) at a cadence of 7~s. The spatial sampling for CRISP and  CHROMIS is 0\farcs058 per pixel and 0\farcs038 per pixel, respectively. This SST dataset was reduced using the SSTRED pipeline 
\citep{2021A&A...653A..68L, % Lofdahl+ SSTRED
Rodriguez2015}. % Jaime+ CRISPRED
High image quality was achieved with the aid of the SST adaptive optics system \citep{Scharmer2024} % Scharmer+ AO
and image restoration with the Multi-Object Multi-Frame Blind Deconvolution \citep[MOMFBD;][]{Noort2005} method. 
The spectropolarimetric \ion{Fe}{i}~6173~\AA\ observations were inverted using a Milne-Eddington inversion code 
\citep{2019A&A...631A.153D} % Jaime ME inversion code
to produce maps of the magnetic field strength along the line of sight ($B_\textrm{LOS}$).

\subsection{Space-borne observations}  
For the extreme ultraviolet (EUV) and magnetic field observations, we used  Atmosphere Imaging Assembly \citep[AIA;][]{Lemen2012} and Helioseismic Magnetic Imager \citep[HMI;][]{Scherrer2012} respectively, onboard the Solar Dynamics Observatory \citep[SDO;][]{Pesnell2012}.
AIA observes the full solar disk which is ideal for studying the large scale jet evolution in coronal diagnostics.
The AR was observed with all EUV (94, 131, 171, 193, 211, 304, and 335 \AA) and UV (1600 and 1700 \AA) channels of AIA with cadence of 12 sec and 24 sec respectively. The pixel size of AIA is 0\farcs6. HMI observes the photospheric magnetic field  with a cadence of 45~s and pixel size 0\farcs5. We use these magnetograms for the AR evolution, starting one day prior to the beginning of the jet activity. In addition to that, we also use the Spaceweather HMI Active Region Patch \citep[SHARP;][]{Bobra2014} data for making the magnetic field extrapolations. This SHARP dataset is available for each AR in the solar disk with a cadence of 12 min. 
To compare ground-based observations with SDO observations, we aligned the AIA and HMI datasets with the SST observations by matching the AIA and HMI spatial sampling to that of the SST.
%by increasing the AIA and HMI resolution to match that of the SST. 
This alignment procedure was carried out using SSWIDL, as explained in \url{https://robrutten.nl/Recipes_IDL.html}.

%--------------------------------------------------------------------
\begin{figure*}[ht!]
   \centering
\includegraphics[width=\textwidth]{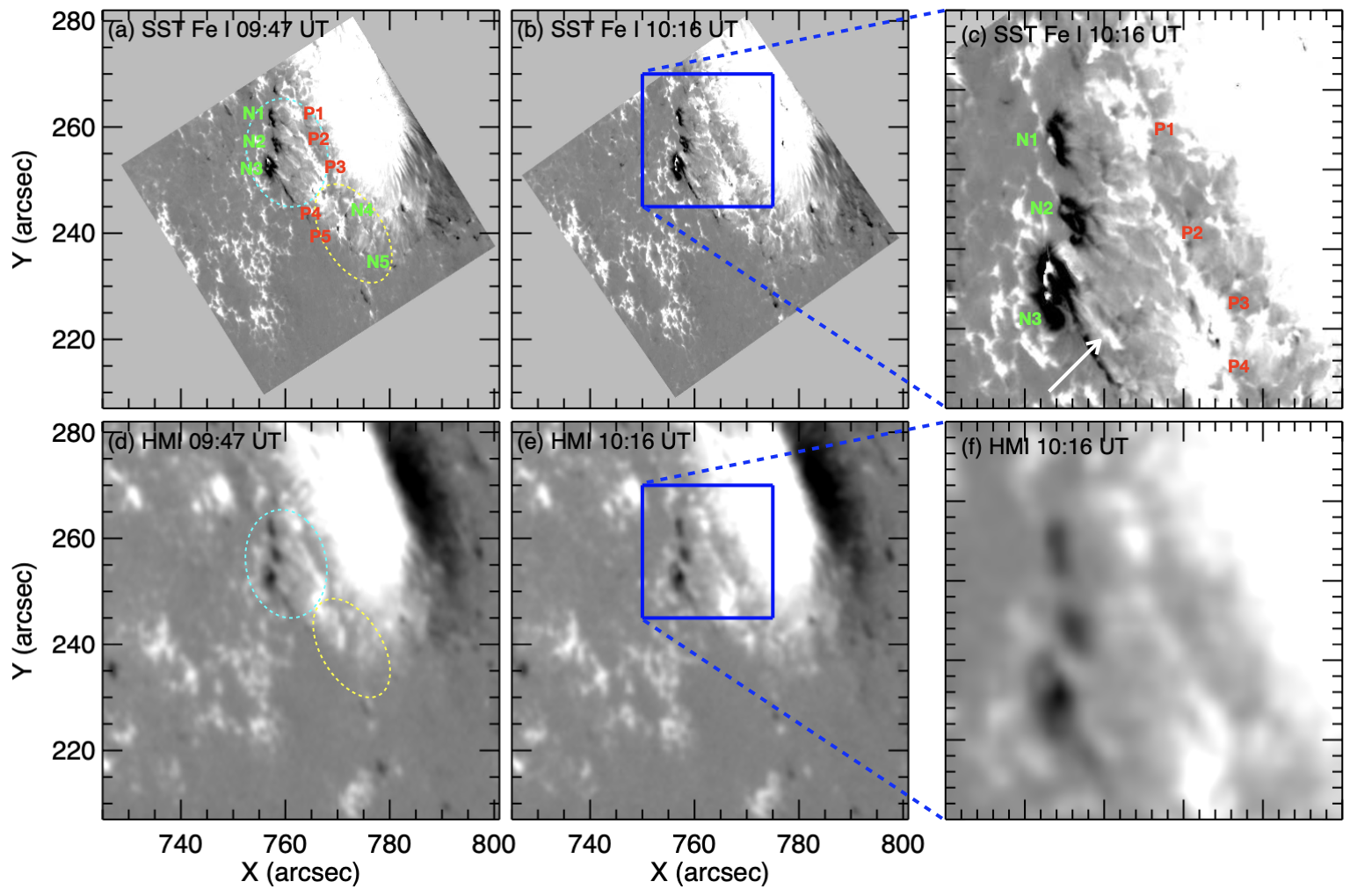}
   \caption{Magnetic field configuration resulting from
   the emerged magnetic flux: left column is  observed with CRISP and right column is observed with HMI.
   Two systems of negative and positive polarities can be identified: in the North N1, N2, N3 with P1, P2, P3 and in the South N4, N5 with P4 and P5. We can see the weak negative polarities N4 and N5 in CRISP observation (panel a). These are moving continuously and the HMI spatial resolution is not sufficient to resolve them. The right column 
   zooms on the blue rectangular area shown in the middle row. The white arrow in panel (c) shows the magnetic polarities moving between the negative (N3) and positive (P3-P4) boundaries observed in the longitudinal magnetic field maps. An animation is attached online with this figure to show the magnetic field evolution from 09:47 to 11:30 UT (see \url{http://tsih3.uio.no/lapalma/subl/afs/fig3_HMI_CRISP.mp4}). The threshold of the magnetic field is +/- 500 G.}
              \label{SST_HMI}%
    \end{figure*}
 \begin{figure*}
   \centering
   \includegraphics[width=\textwidth]{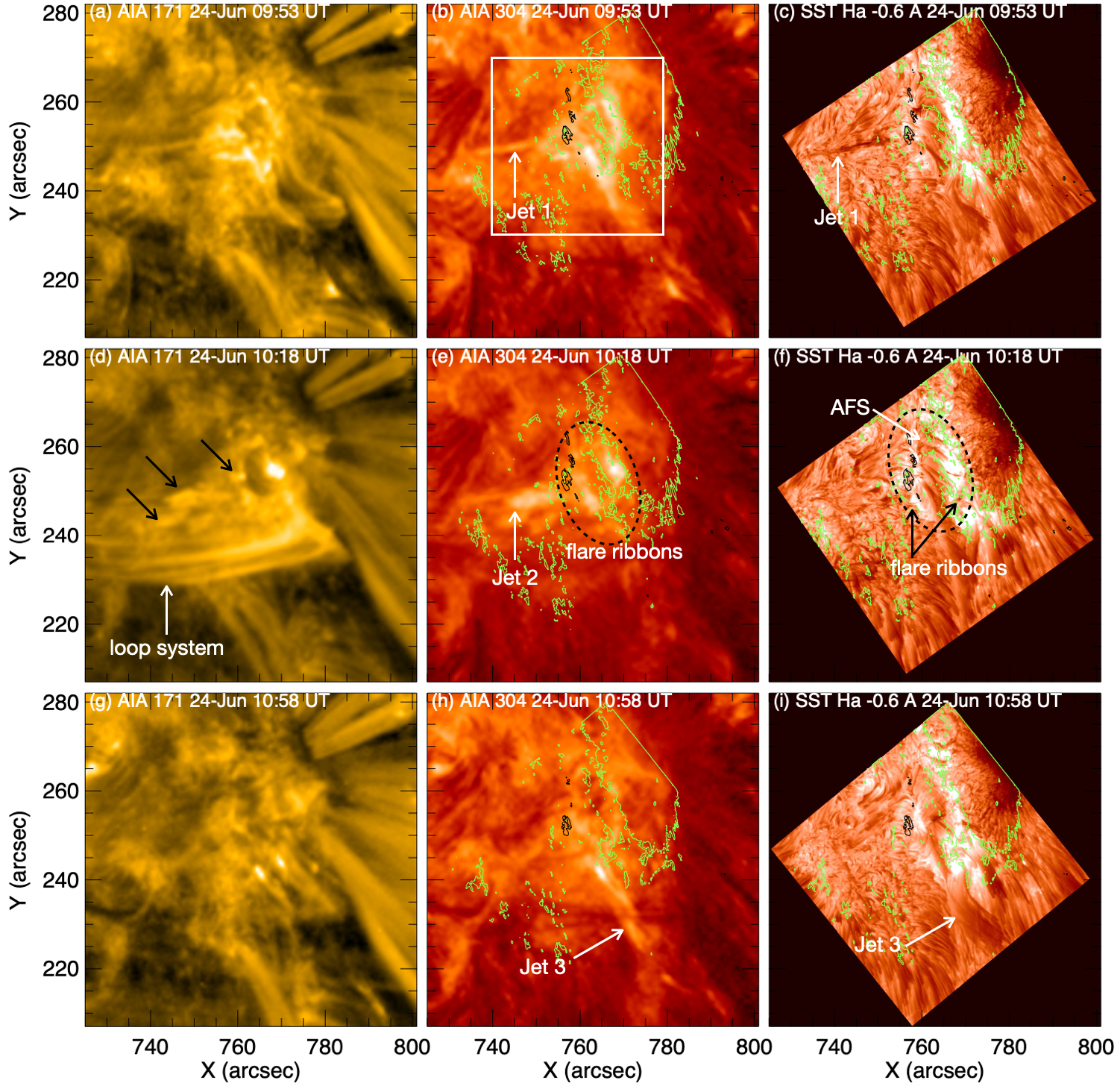}
   \caption{EUV (171 and 304 \AA) and H$\alpha$ observation of jets at their maximum height. Magnetic field contours of strength $\pm $ 500 G from SST/CRISP  \ion{Fe}{i}~6173~\AA\  are over-plotted in second and third column for all three jets. Jet 1, Jet 2, and Jet 3 are shown with white arrows and the ellipsoidal shaped flare ribbons around the AFS are shown with black dashed contour in panels (e-f) and the ejected blobs of Jet 2  are indicated by black arrows in panel (d).  The bright loops (West to East) heated by the flare are shown in panel (d). The white rectangular box in panel (b) shows the FOV for Fig.~\ref{fig:afs}. 
   }
    \label{SST_AIA}%
    \end{figure*}
%figure5
\begin{figure*}
\centering
   \includegraphics[width=\textwidth]{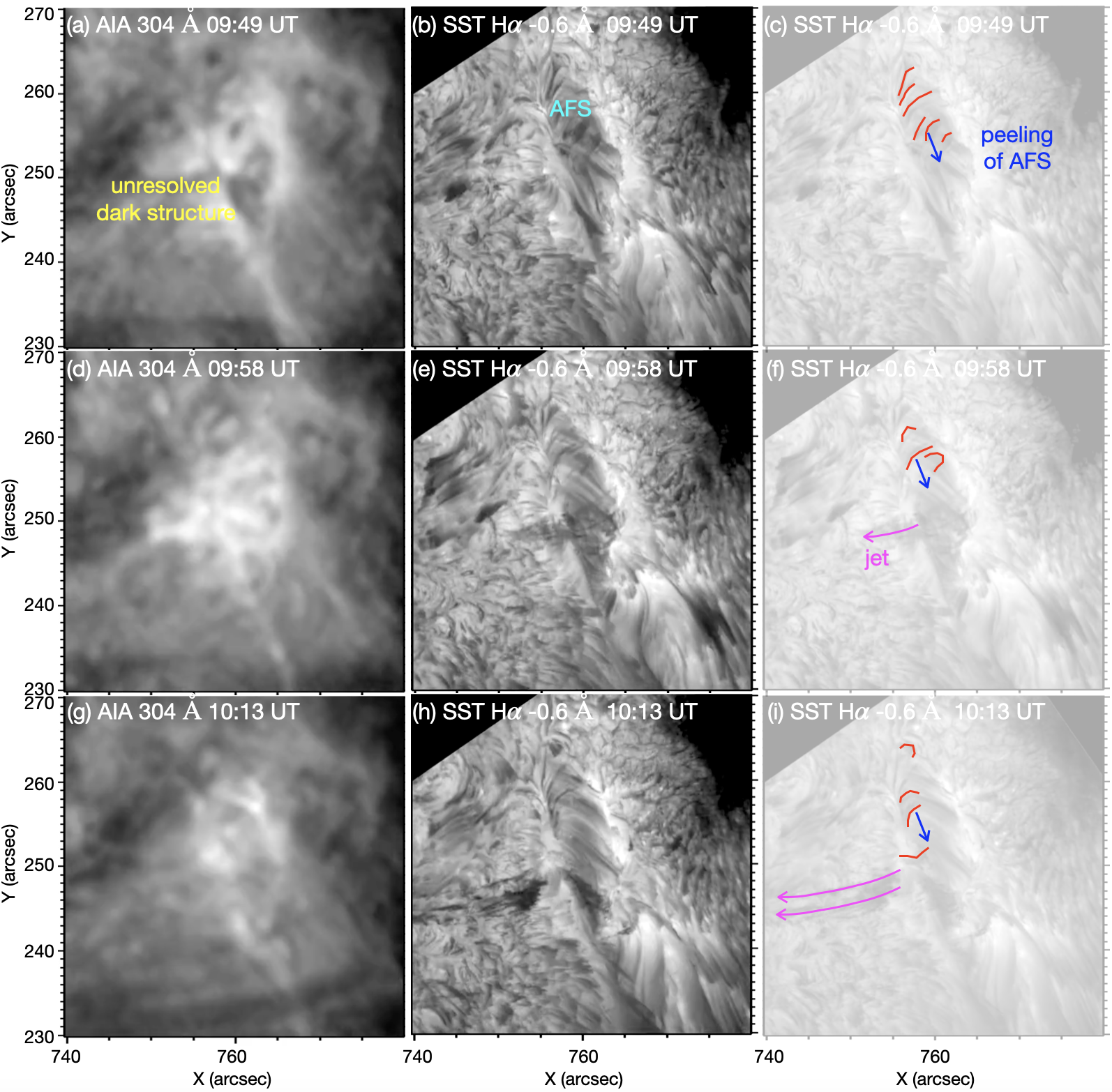}
   \caption{A zoomed-in view of the AFS region, corresponding to the field of view displayed in Fig.~\ref{SST_AIA} (panel b), is presented. The left column displays observations in the AIA 304 \AA\ wavelength, the middle column shows SST observations in the H$\alpha$ blue wing, and the right column is the same as the middle but in pale rendering. The right column was used to manually trace the peeled stripes (marked in red) from the AFS. The blue arrow indicates the direction of the peeling material flowing southward and contributing to the jet (highlighted in magenta color). The feature observed in AIA 304 \AA\ appears as an unresolved dark structure, which is resolved in SST observations as an AFS. This figure is accompanied by an online animation (see \url{http://tsih3.uio.no/lapalma/subl/afs/fig5_AFSpeeling.mp4}).
   }
    \label{fig:afs}%
    \end{figure*}
  %Figure 5     
\begin{figure*}
   \centering
  \includegraphics[width=\textwidth]{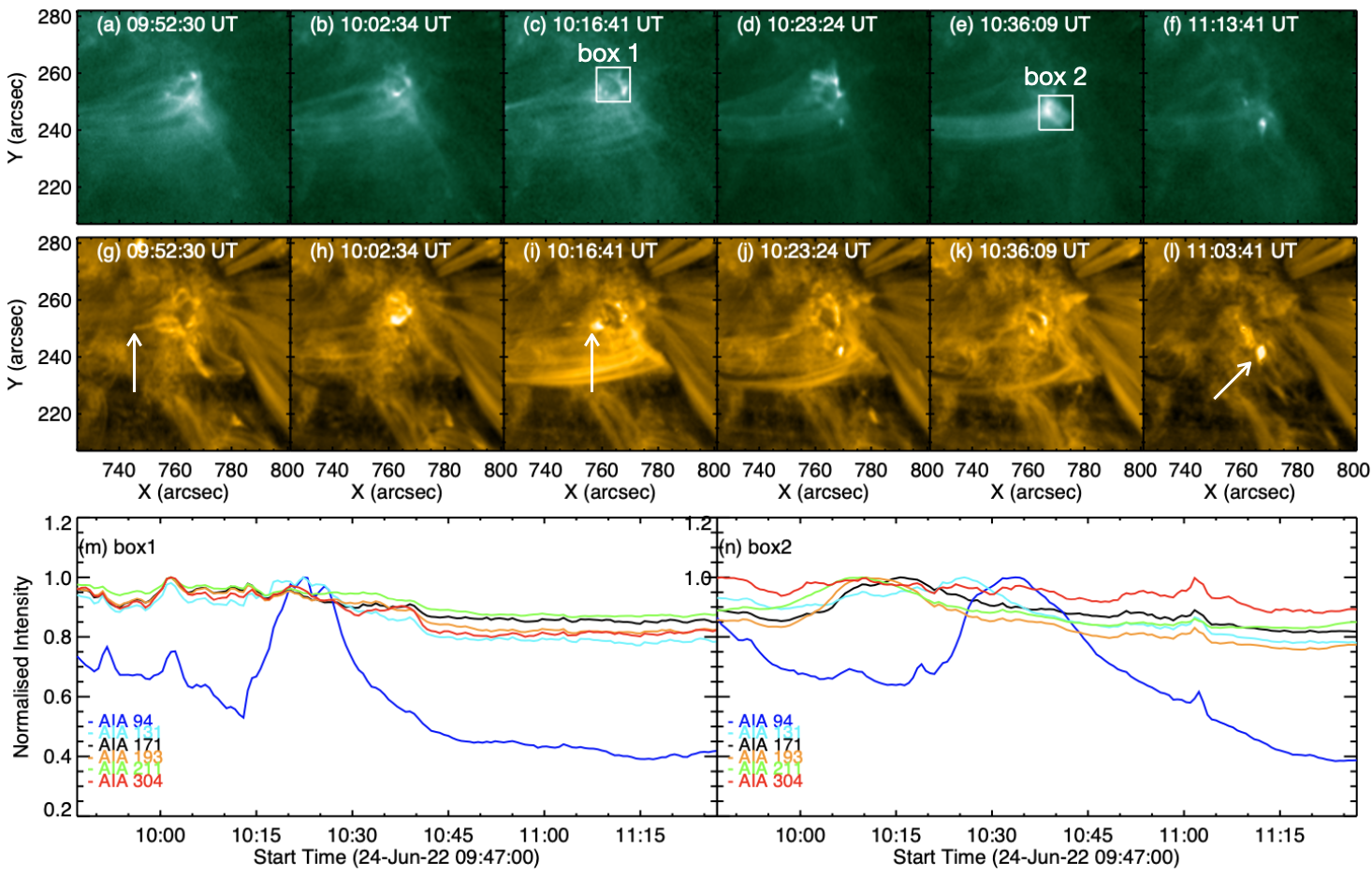}
   \caption{Drifting of the brightening from North (box1) to South (box2). Top row: AIA 94 \AA\, middle row: AIA 171 \AA\, bottom row: temporal evolution of the normalised intensity in six AIA channels in box1 and box2. Three jets are marked with white arrows in panel g, i, and l respectively. 
   Drifting of the hot material is quantified  in the hot AIA 94 \AA\ channel with the normalised intensity curves (blue curve in the bottom row).}
   \label{slipping}
    \end{figure*}
    
%Figure 6 
    
\begin{figure*}
   \centering
   \includegraphics[width=\textwidth]{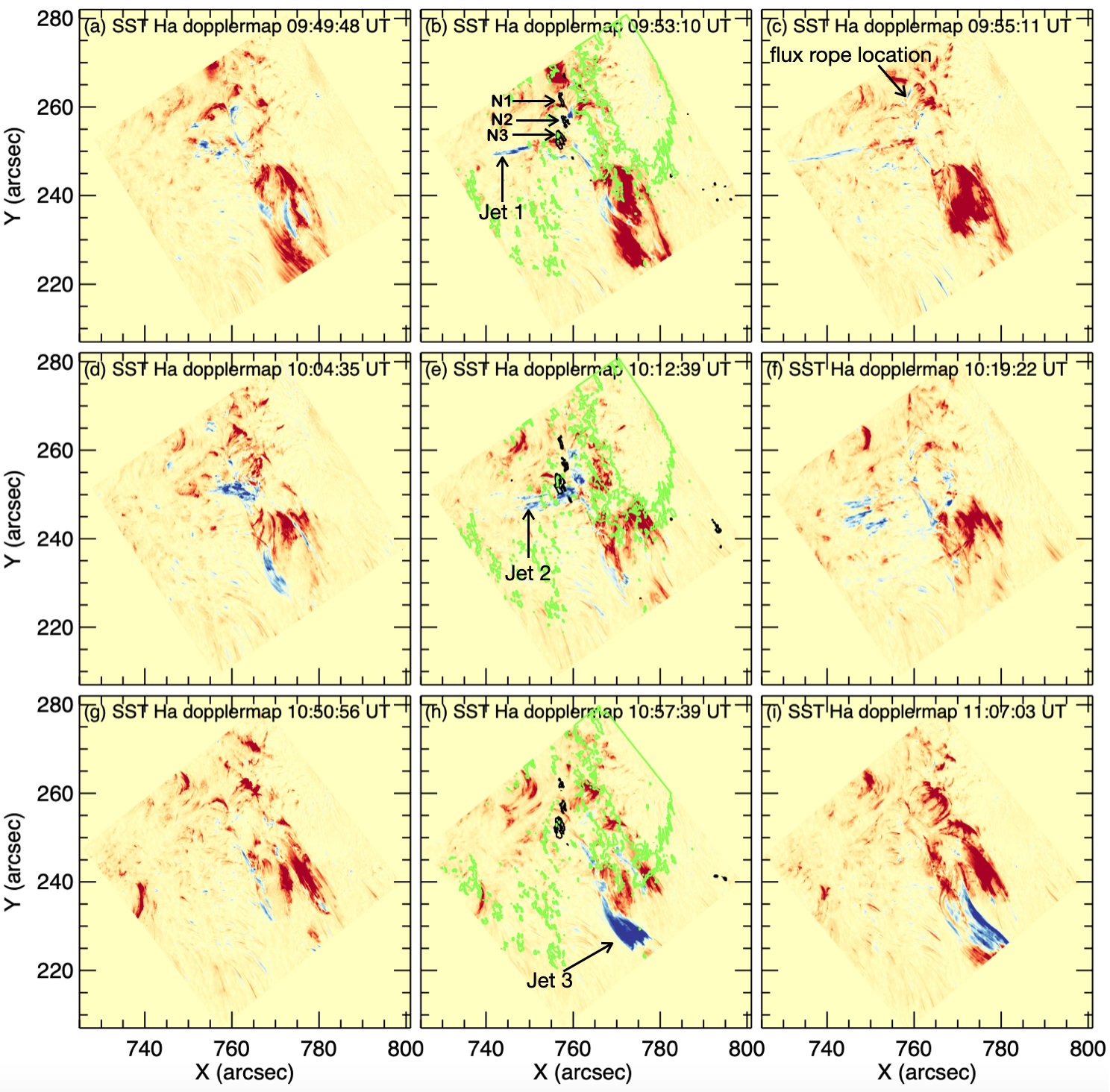}
   \caption{SST/CRISP H$\alpha$ Dopplermaps showing the evolution of the three jets. The location of the dynamic swirling flux rope is shown in panel c (see the attached animation). These Dopplermaps are constructed using two H$\alpha$ line positions offset by 1~\AA\ from the line center ($\pm$45~km~s$^{-1}$).
   Magnetic field contours of $\pm$500~G (black and green) from the SST \ion{Fe}{i}~6173~\AA\ magnetograms are overlaid in the first column. 
   This figure is associated with an online animation (see \url{http://tsih3.uio.no/lapalma/subl/afs/fig7_dopplermaps.mp4}).
  }
   \label{doppler_maps}
    \end{figure*}

\begin{figure*}
   \centering
   \includegraphics[width=\textwidth]{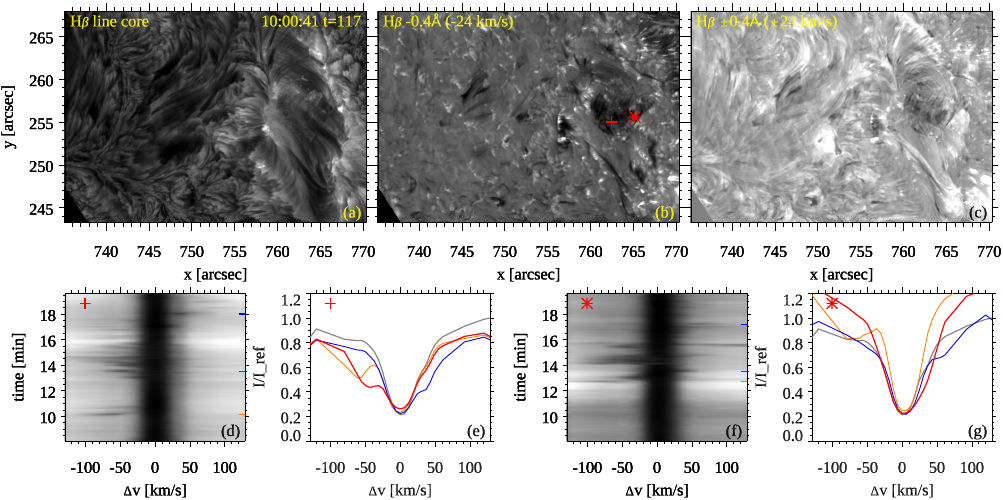} \\
   \includegraphics[width=\textwidth]{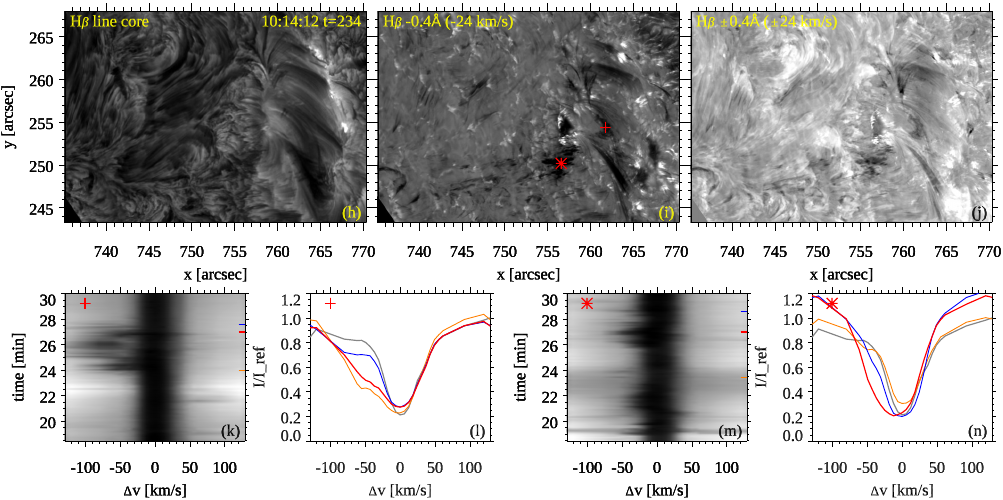}
   \caption{ 
   CHROMIS H$\beta$ observations of the swirling flux rope (panels a--g) and Jet 2 (panels h--n). For two time steps (10:00 UT and 10:14 UT), the row of images (a--c, and h--j) show H$\beta$ line core, blue wing, and a Dopplergram, respectively. In the H$\beta$ wing images, a red plus and an asterisk symbol mark the locations for which the spectral evolution in $\lambda t$ diagrams are shown in panels (d), (f), (k), and (m). The spectral profiles for these locations for the two chosen time steps are shown in red in the profile panels (e), (g), (l), and (n). For each $\lambda t$ diagram, two more spectral profiles from different times are shown in orange and blue. The times for these profiles are indicated with short colored dashes in the $\lambda t$ diagrams. The thin grey profile is a reference H$\beta$ profile that is the average over a quiet region. The time along the $t$ axis in the $\lambda t$ diagrams is in min after the start of the observation at 09:47:08 UT. An animation of the three image panels is provided in the online material (see \url{http://tsih3.uio.no/lapalma/subl/afs/fig8_hbcore+hbwing+dop.mp4}).}
                 \label{Chromis}%
    \end{figure*}

\begin{figure}[ht!]
   \centering
   \includegraphics[width=0.5\textwidth]{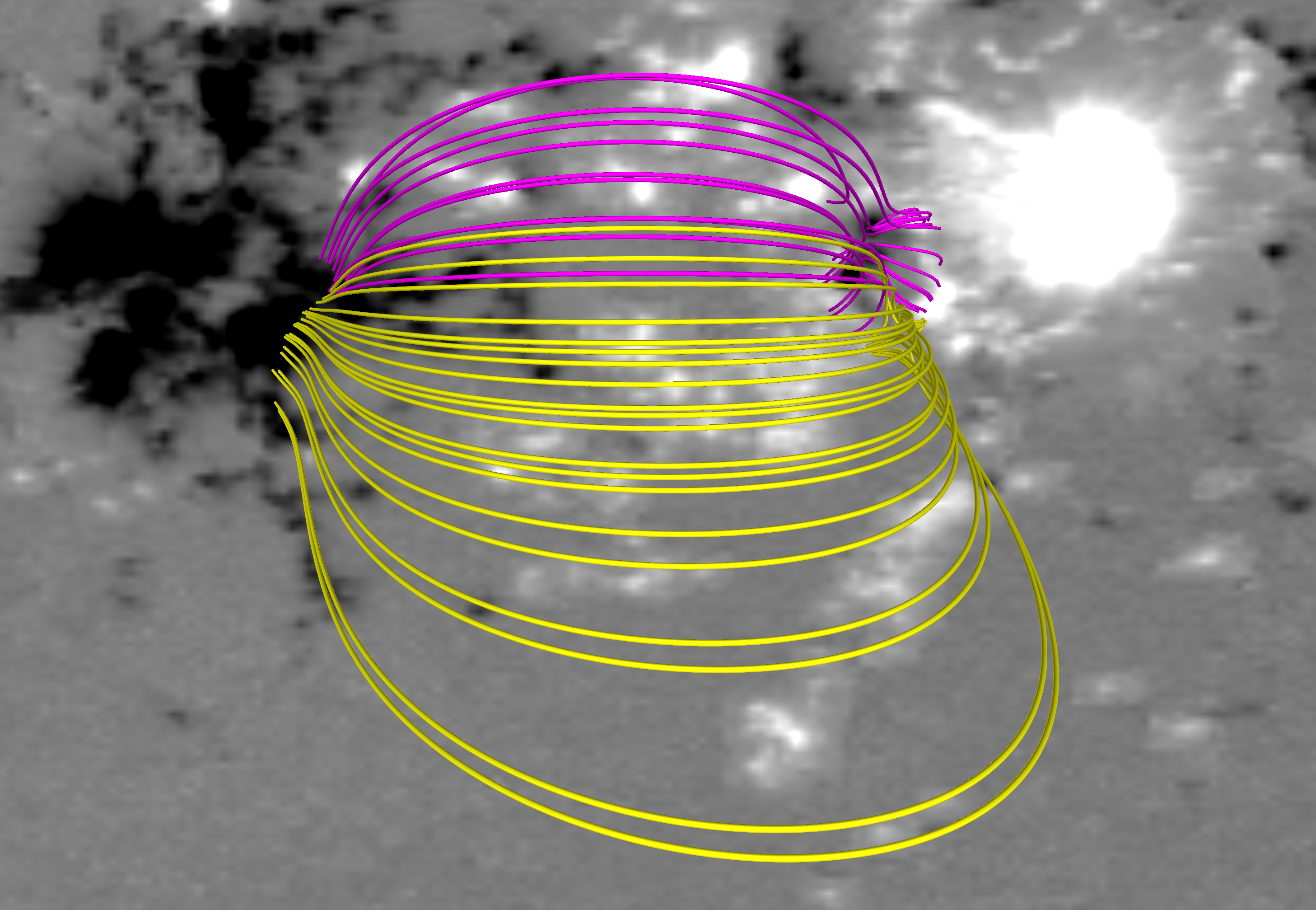}
     \caption{Non-Force Free Field (NFFF) extrapolation of AR 13038 (top view) at 10:12 UT. This rendering shows selected magnetic field lines with magenta field lines
       connecting to the region containing the N1, N2, N3 with P1, P2, P3
     polarities. 
The yellow extrapolated field lines show the field line geometry 
in the southern part of the AR and are co-spatial with the ejection of Jet~3.
   }
              \label{exp1}
    \end{figure}

\begin{figure}
   \centering
   \includegraphics[width=0.5\textwidth]{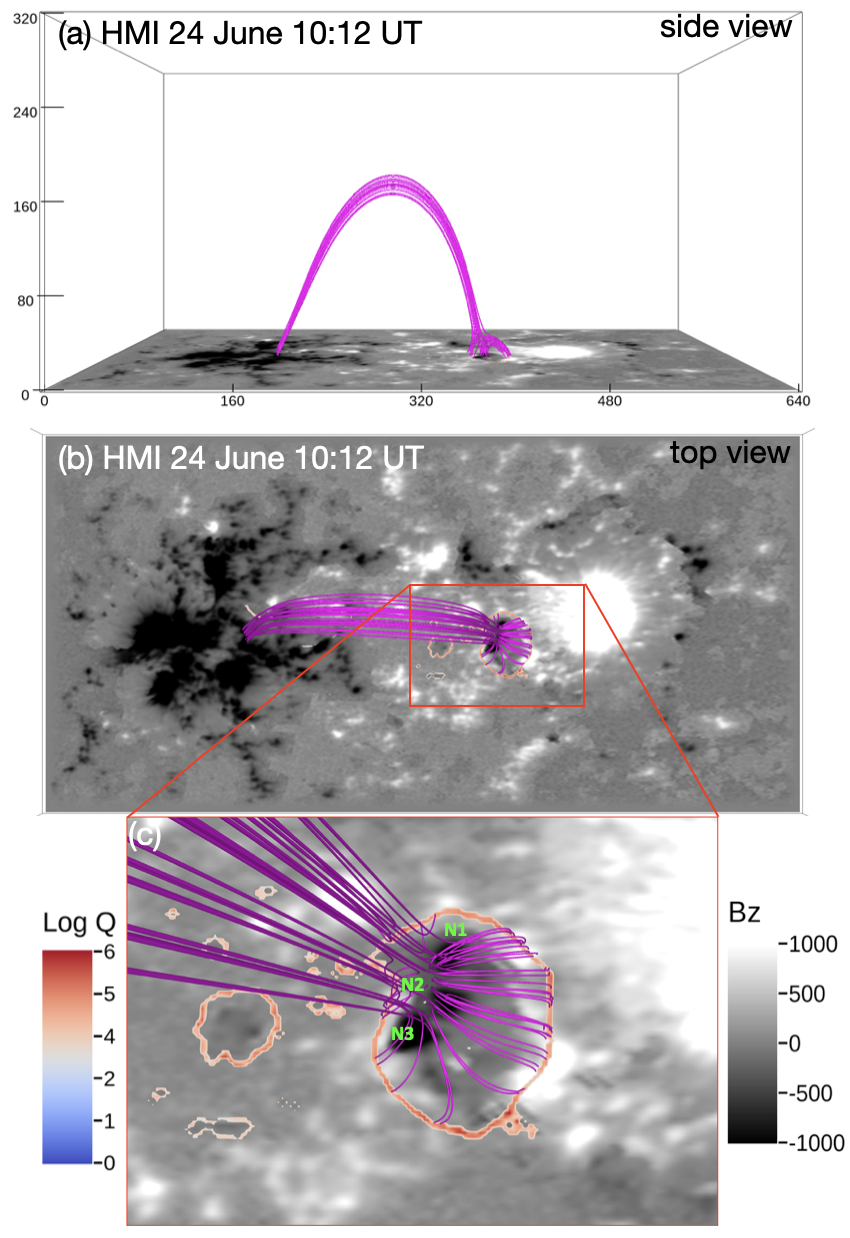}
   \caption{Different views on the NFFF extrapolation: side view (panel a), top view (panel b), zoom FOV of top view (panel c). The three negative polarities N1, N2, and N3 are connected to the positive polarities P1, P2, P3 nearby with small inner spines, and globally to the plage region (large negative polarity region on the East). The squashing factor Q is plotted at the photospheric level in panels b and c (colored contours) to show the separatrix at the periphery of the fan surface around N1, N2, and N3.}
              \label{exp2}
    \end{figure}

\begin{figure}
   \centering
   \includegraphics[width=0.5\textwidth]{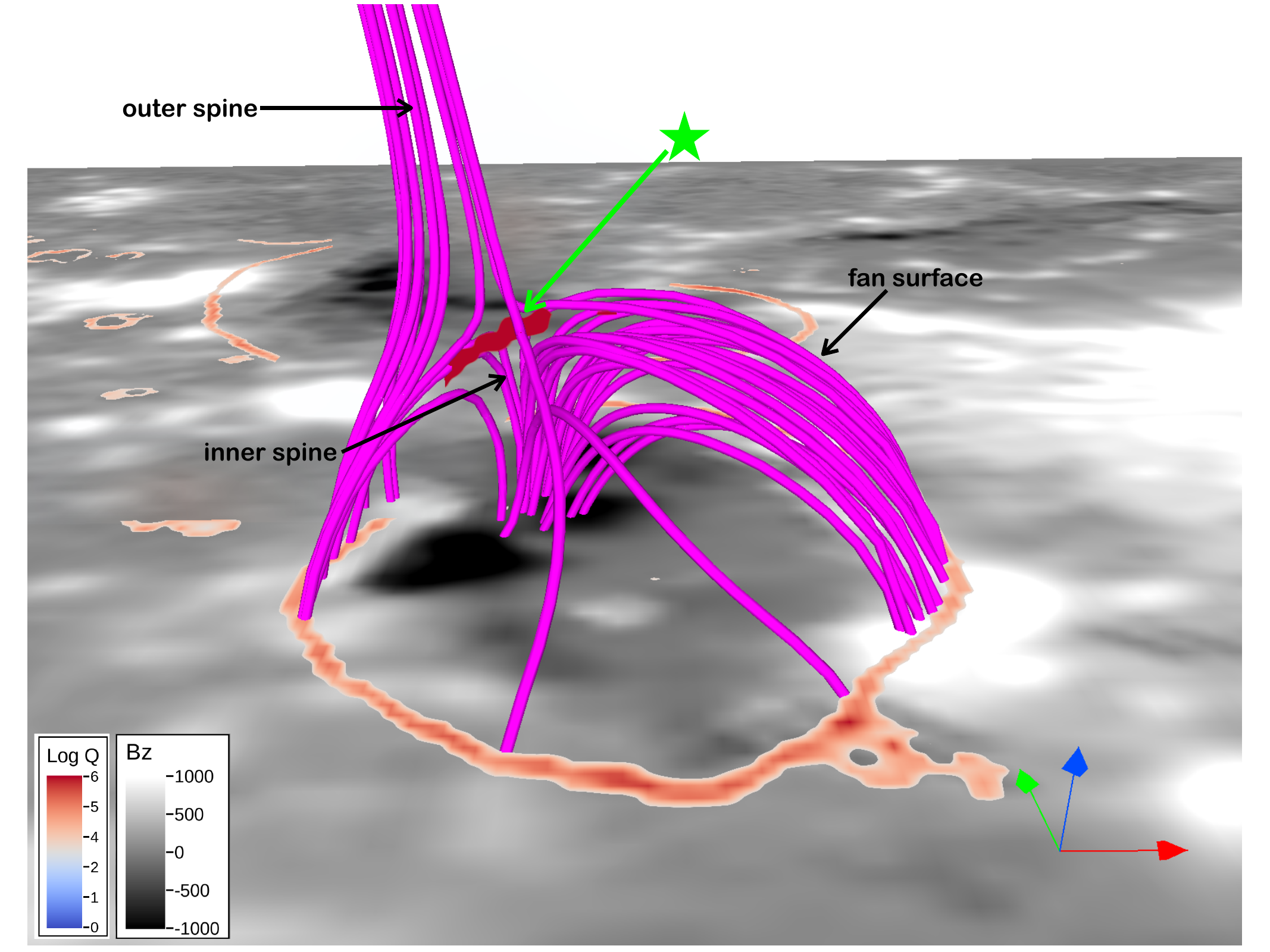}
   \caption{
   Side view of the fan-spine topology with a clear view on the location of the null point. The null point location at the top of the fan surface (the site of the widest high-Q contour indicated with an elongated red surface) at a height of 6~Mm above the photosphere is marked with a green arrow and star. 
   The approximate location of the fan surface, the inner spine (linking the null point to the parasitic negative polarities), and the outer spine (moving away from the null point) are shown with black arrows.} 
              \label{NFFF}%
    \end{figure}

  \section{Recurrent jet environment}
  \label{sec:recurrent}
On 24 June 2022,  AR 13038 was the most jet-producing region in the Sun with several GOES B and C class solar flares. We focus on  the jets  and their environment occurring during the  SST observations from 09:47 UT to 10:55 UT . 

 \subsection{Active region evolution and magnetic flux emergence}
 \label{sec:magnetic_flux_emergence}
 Figure \ref{AIA_HMI}  presents a global picture of AR 13038 located near the West limb (N 15, W60).
 A part of this AR was observed with SST which is shown in the last row of Fig.~\ref{AIA_HMI}, with a rotated white square. 
 The basic magnetic configuration of this region consists of a positive polarity leading sunspot and an extended negative polarity region trailing on the East side. The negative polarity patch on the upper West side (or limb-side) of the leading positive polarity sunspot is due to the projection effect from its close proximity to the limb. 
 Long loops are observed in coronal temperatures (AIA 171~\AA), joining the two main polarities in the East-West direction in the southern end of the sunspot. 
 Towards the southern direction of the AR  nearly open field lines (long open loops) are observed. 
 On the East side of the sunspot, an extensive region with EMF is found (Fig.~\ref{FigHMI_SST}). The large positive polarity is surrounded by small negative polarities (see the cyan oval) which were formed due to flux emergence over time. 
 Looking at the evolution of magnetic field 
 over the duration of 12~h, from June 23 22:00~UT to June 24 11:00~UT, the small negative polarities are moving to the North-East indicating a strong shear between the negative and positive magnetic environment. The motion of the small negative polarities stops when reaching the former positive magnetic network located between the two main polarities. The zoom image of this area with SST/CRISP (Fig.~\ref{SST_HMI} panels c and f) shows more details of the small negative polarity patches.

 Two systems can be identified with labels N1, N2, N3 along with P1, P2, P3 encircled in the cyan oval, and N4, N5 along with P4, P5 in the yellow oval. The latter small polarities patches are not detectable in the HMI magnetograms. 
 As a consequence of the continuous magnetic flux emergence for 
 an extended time, the motion of small new magnetic polarities has been observed in the longitudinal magnetic field maps, between the negative polarities (N1, N2, N3) and the positive polarities (P1, P2, P3). Due to the location of the AR towards the limb, the longitudinal magnetic field measurements have a strong horizontal component. Therefore, horizontal fibril-like structures between small positive and negative polarities are visible (an example is shown with a white arrow in Fig.~\ref{SST_HMI} panel c).
 Their direction  is inclined by an angle of about  $\sim$ 120$^\circ$ with respect to 
 the PIL, indicating a strong shear. 

\subsection{Overview of the jets }
We observed the AR for a duration of $\sim$4 hours (from 08:00--12:00 UT) with SST. The observations by CRISP in H$\alpha$ focus on 
the EMF region  in two  successive time sequences. 
A GOES C1.5 class flare started at 08:49 UT, peaked at 08:58 UT, and decayed at 09:13 UT, and was followed by a jet towards the South at 09:17 UT and a jet towards the East at 09:32 UT. Soon after, at 09:59 UT, a GOES  B8.8 class flare started to occur, peaked at 10:20 UT and decayed at 10:30 UT in the GOES record. In the present paper we are focused on the later time series observations during 09:47--10:55 UT. Continuous ejection of jets at 09:53 UT, 10:12 UT, and 10:57 UT was observed associated with small flaring activity. Emergence of magnetic flux was a continuous process for many hours before the activity. New flux  emergence creates favorable conditions for surges, jets and flares to occur during this long process and we observed jets, surges and bright kernels issued from a flare ribbon with ellipsoidal shape (encircled in Fig.\ref{SST_AIA} panel e-f) around the flux emergence region at multiple times (see the attached movie associated with Fig.~\ref{AIA_HMI}). 

In  Fig.~\ref{SST_AIA} we present the jets  observed in AIA 171~\AA, 304~\AA, and H$\alpha$ (Jet 1, Jet2, Jet3). Jet 1 and Jet 2 are going to the East, while Jet 3 is going to the South. After Jet 1 at 09:47 UT, which is apparent in the EUV filters as well as in the blue wing of H$\alpha$, we observed the formation of AFS in the H$\alpha$ line core (see movie  attached with Fig.~\ref{AIA_HMI}).
 We believe that, during the flux emergence process, dense material is lifted up forming a system of AFS \citep[as explained in][]{Schmieder2010, Ma2015}. 
Comparing the images in 171~\AA, in 304~\AA\ and in H$\alpha$ (Fig.~\ref{SST_AIA}), we observed  that  the dark area surrounded by brightenings corresponds to the AFS region between the two flare ribbons in H$\alpha$. The AFS is well visible in the H$\alpha$ images (see  panel f) as dark strands at the top of the ellipsoid covering the flare region. In H$\alpha$, we observe what seems to be a progressively peeling-like mechanism of the AFS followed by the ejection of elongated surges (see Fig.~\ref{fig:afs}). 
In SST observations, we detected numerous stripes of dark material peeling away from the AFS just before the jet eruption. As the jet begins, fewer stripes are visible. From the animation associated with Fig.~\ref{fig:afs}), we can follow the peeled material which is moving in the south direction (see the blue arrow in Fig.~\ref{fig:afs}) last column) and going towards East in the jet. 
This resembles the peeling mechanism of the external layers of the emerged domes, frequently reported in flux emergence experiments, that precedes the surge ejection 
\citep[see, e.g., the animation of Figure 10 by][]{Moreno2013}. In the AIA filters, the dense and cool material absorbs a part of the EUV emission and appears as an unresolved dark structure (see Fig.~\ref{fig:afs}) panel a), which could have been  interpreted as a mini filament eruption, in the absence of high resolution SST observation.
%\citep[see, e.g.,][and references therein]{Muglach2021}.

There are many blobs being ejected alongside the Jet2 in AIA 171 and 304 \AA\ (shown with the black arrows in Fig.~\ref{SST_AIA} panel d). Another arcade occurs for a second time  after Jet 2 between 10:22 UT and 11:00 UT  covering the former AFS with a different direction (NW to SE). They could correspond to  post flare loops after  reconnection and   cooling  hotter loops visible in AIA 171 \AA. Flare ribbon brightenings are apparent 
in Fig.~\ref{SST_AIA} in the negative and positive magnetic field.  
Around 10:18 UT (Jet 2), a set of East-West bright loops  are observed in AIA 171 \AA\ (panel d). The West footpoints of the bright loops are identified as a  North-South (NS) elongated feature. 
This NS feature  is co-spatial with the positive polarities P3, P4, P5 included in the yellow  oval with N4 and N5 in  Fig.~\ref{FigHMI_SST} panel a. 
Jet 3 appears thin shaped in 171 \AA\ and 304 \AA\ (Fig.~\ref{SST_AIA}, panels g, h) and corresponds to a broad dark jet in H$\alpha$ (panel i).

High temperature brightenings are detected in the hot AIA channels (the best is seen in the AIA 94 \AA\ channel in Fig.~\ref{slipping}). 
The light curves obtained in two selected boxes (marked in Fig. \ref{slipping} first row) indicate the normalised intensity at different times (Fig.~\ref{slipping} bottom row). The total intensity has been normalised with the maximum value of the intensity of the time series. 
These two boxes are of size $\sim$10$\arcsec$. 
The intensity variation at the different box locations 
in 6 EUV channels of AIA are plotted in different colors.
Box1 in the North of the EMF reaches maximum intensity between 10:20--10:25 UT, while box2 in the South reaches its maximum between 10:30--10:40 UT. These brightenings are associated with the jets towards the East at 09:53 and 10:18 UT and towards the South at 10:58 UT; 
 they only show a clear intensity peak in the hot AIA 94 \AA\ channel. We believe that the plasma in the brightening gets such high temperature that it is more clearly visible in the hot AIA 94 \AA\ wavelength. 
 The brightening bounded by box1 in the North is observed to move towards the South (see box2) in $\sim$ 18 minutes.  
 
This is a clear indication that the reconnection site is moving from North to South and the overlying magnetic field lines have different directions  to the East or South that explains the change of direction of the jets.
The distance between the two locations of the brightening in Fig. \ref{slipping} (c and e) is estimated to be 20 Mm. 

\subsection{Dopplershifts in H$\alpha$}
\label{sec:doppler}
On the basis of the spectroscopic observations from CRISP, one can calculate Doppler shifts using the H$\alpha$ lines. We have constructed Dopplergrams by subtracting two different H$\alpha$ line positions at $\pm$45~\kms\  offset from line center. In Fig.~\ref{doppler_maps} we are showing the resulting color maps for the vertical velocity component, with the downflows drawn in red and the upflows in blue, at different times from 9:49:48  UT (top left panel) to 11:07:03 UT (bottom right panel). 
Negative and positive magnetic field contours ($\pm$500 G) are plotted in the second column (for all three jets) in black and green, respectively, using the \ion{Fe}{I}~6173~\AA\ magnetograms from SST. Here, upflowing jet material is visible between the negative polarities and the positive polarities x=760$\arcsec$--770$\arcsec$, y=230$\arcsec$--240$\arcsec$ (see panels b,e, and h). The maps clearly show the jet flows as elongated blue features (and blobs, in the case of Jet 2); interestingly, one can also identify a perturbation that can be interpreted as a moving flux rope evolving from North to South. The initial location of this perturbation is shown with a black arrow in panel c; the motion can be discerned in the associated movie as blue-shifted plasma starting in between the N1, N2, N3 and P1, P2, P3 region and moving toward the South, with subsequent ejection of plasma material as Jet 2 at $\sim$10:05 UT. That this motion can be interpreted as a flux rope (in fact, with right-handed twist) will be shown in the following section (Sec ~\ref{sec:chromis}), with the help of H$\beta$ observations.
  
The motion of the flux rope also causes collimated downflows oriented towards the South (panel e, f).
For the southern region from P4, and P5, from 09:49 UT until 10:12 UT, there are mainly downflows (red shifted material) around x=770\arcsec--780\arcsec\ and y=220\arcsec--240\arcsec.
Then from 10:57 UT (Jet 3) there are upflows (blue shifted material) and Jet3 continues till 11:07 UT  with up- and  downflows.  The base of Jet3 is approximately close to the negative polarity N3 plotted in black contours. The downflows of Jet 3 are observed in many parallel structures covering a large area. The footpoints of these structures extend  $\sim$ 15$\arcsec$ towards south in the region of P5 (green contours in the last row).

\subsection{Swirling flux rope in H$\beta$ observations}
\label{sec:chromis}
Inspecting the high cadence (7~s) CHROMIS H$\beta$ 4861~\AA\ observations in different spectral line positions, similar to H$\alpha$ Dopplermaps, we found evidence of a cool plasma cloud moving from the North of the EMF region (above the negative polarity N1) and ejecting the plasma material as Jet 2 towards the East from the location of negative polarities N2, and N3 (see Fig.~\ref{Chromis} and the associated animation); 
we believe that this twisting material corresponds to the flux rope swirling over the AFS region and flowing away towards the East discussed in Sec.~\ref{sec:doppler}. 
To test  the twisting motion of the flux rope, we performed  spectral analysis of the H$\beta$ profiles in the AFS region, presented in Fig.~\ref{Chromis}. We chose two time steps at 10:00 UT and 10:14 UT, shown in the upper (panels a--c) and lower half ((panels h--j)) of Fig.~\ref{Chromis}, respectively;  we selected two different points in each (shown with plus and asterisk symbols in red in the respective $H\beta$~$-0.4$~\AA~panels) to calculate the spectral evolution by means of $\lambda$ -- t  diagrams. The spectral profiles at the flux rope location are plotted for various time steps. These specific time steps are indicated by blue, red, and orange dashes on the right axis of the spectra (see panels d, f, k, and m). The spectral profiles for different locations within the flux rope, marked by red asterisks and plus signs, are shown in panels e, g, l, and n. Exploring the spectra and profiles at different times, it is interesting to follow the dynamics of the plasma at the flux rope location, as we see clear asymmetries and even separate components in both wings. For example, the red and orange profiles in panel e have a separated component at a Doppler shift of about $-50$ and $-60$~km~s$^{-1}$. The blue profile in this panel e has a red shifted component at about $+40$~km~s$^{-1}$. Panel g also shows examples from this region with similar Doppler shifted components. 
The associated animation shows the temporal evolution of the AFS region in the H$\beta$ line core (left panel), blue wing ($-0.4$~\AA, central panel), and Doppler maps ($\pm$0.4~\AA, right panel) in the interval 09:47 UT to 10:33 UT. 
From 09:55 UT, we see highly dynamical small dark threads that appear to move southward and someof them in a swirling pattern above the AFS region. These dynamical threads or stripes are shown in Fig.~\ref{fig:afs} in red color, explained as they are peeled out from the AFS. The swirling pattern of a big dark plasma cloud is particularly clear in the blue wing and Doppler movie and we interpret this as a swirling flux rope (similarly seen in H$\alpha$ Doppler map observations). This swirling flux rope travels southward over this region and evolves into the ejection of Jet 2 towards the East at 10:06 UT.

We conjecture that these rapidly changing blue and red shifts in the spectra correspond to the twisted flux rope passing from North over the AFS and contributes to the Jet 2 material. 
At 10:00 UT, the spectral profiles at different times at two locations (see the asterisk and plus sign in panel b) show a blue and red shift (panels e and g), where the spectral profiles have the red and blue asymmetry in comparison to the average profile at a quiet region (grey color).
At the time of Jet 2 at 10:14 UT, the profiles show a clear blue shift of more than $-60$~km~s$^{-1}$ (see the blue curve in panel l corresponding to $t=$27~min in panel k). Similar blue shifts are also evident in other profiles from the jet material (location of the asterisk in panel i and spectra in panel m and n).
In this way, these very highly resolved observations with a higher cadence (as the H$\alpha$ observations has a cadence of 40 s in comparison to H$\beta$ with 7 s)  
are showing the highly dynamical swirling motion of the flux rope over the AFS with alternate blue and red shifts, and only blue shifts in the jet.  

\section{Non-force-free magnetic field extrapolation}
\label{sec:NFFF}
\subsection{Method}
We apply the  Non-Force-Free Field  (NFFF) extrapolation technique, as developed by \citet{Hu2008a} and \citet{Hu2010}, to extrapolate the coronal magnetic field of AR 13038. The time-frames under investigation are from 10:00 to 11:12 UT on June 24, 2022, which include the time of the vector magnetograms portrayed in Figs.~\ref{exp1} and~\ref{exp2}.

 The NFFF extrapolation technique is built on the minimum energy dissipation rate principle (MDR), which assumes that the plasma system tends towards a state with the lowest possible dissipation rate on a time-scale shorter than the dynamical one. In that case, the magnetic field, denoted as $\mathbf{B}$, fulfils the double-curl Beltrami equation \citep{Bhattacharyya2007}. The field $\mathbf{B}$ is consequently described as the sum of two linear-force-free fields and a potential field \citep{Hu2010}. This results in a magnetic field configuration which can have non-zero Lorentz force.
An iterative method is then applied to the NFFF extrapolation that minimises the average discrepancy between the observed and calculated horizontal field at the photospheric boundary. The NFFF technique has recently been effectively used to interpret various transient phenomena in active regions, such as flares, coronal jets, and coronal dimmings \citep{Prasad2018, Nayak2019, Prasad2020, Prasad2023}. 
We recognize the unreliability of vector magnetograms when the region of interest is too far from the disk center, as explained in \citet{Bobra2014, Falconer2016}. In our case the region is at 58$^\circ$, so, to miminize the problem, we employed SHARP data remapped from CCD coordinates to a heliographic Cylindrical Equal-Area (CEA) projection centered on the selected  patch for the NFFF extrapolation. We also visually compared the SHARP data presented in Fig.~\ref{exp1} with the SST magnetograms and did not find any projection artifacts, such as the fictitious negative polarity on the limb side in the $B_\textrm{LOS}$ maps in Fig.~\ref{FigHMI_SST}. The magnetogram shown in Fig.~\ref{exp2} panel (a) spans $640\times 320$ pixels in the $x$ and $y$ directions of a Cartesian coordinate system, while the vertical extent of the domain is set to be $320$ pixels. In physical terms, the dimensions of the computational domain are approximately $466$ Mm $\times$ $233$ Mm $\times$ $233$ Mm in the $x$, $y$, and $z$ directions, respectively.
 
\subsection{Magnetic topology}

The NFFF extrapolation using HMI vector magnetic field maps shows 
 the existence of a null point over the region of negative polarities N1, N2, N3 connecting with the positive polarities on the West side (Fig.~\ref{exp2}). The second system of negative polarities (N4, N5) is not very prominent in the HMI resolution, so it is not possible to look for a second null point in that location using extrapolations from the  HMI data. For extra guidance, we also calculated
 maps of the squashing factor $Q$ 
\citep{Priest_Demoulin_1995, Demoulin_etal_1996, Titov_Hornig_2002, Liu2016} on the
photospheric plane. 
Narrow regions with very large  $Q$ values (plotted with colored contours in panels b and c of Fig.~\ref{exp2}) indicate a strong gradient in the remote connectivity of the field lines issuing from them, i.e., they mark clear candidates for the photospheric base of a separatrix or quasi-separatrix layer (QSL).

The result of the extrapolation is particularly interesting for the
  field lines starting in the region around the N1, N2, N3 negative
  polarities (Fig.~\ref{exp2}). Those negative magnetic concentrations act as
  parasitic polarities surrounded by positive patches (see, e.g., Fig.~\ref{exp2},
  panels b and c or Fig.~\ref{exp1}); consequently, the presence of a null point above  them is a natural feature of the magnetic configuration. In
  the figure, the isocontours for large values of the squashing factor $Q$, drawn on top of
  the photospheric magnetogram, show a simple closed-circuit shape; the field lines that leave the photosphere at the site of the
  widest high-$Q$ contour are seen to delineate a fan surface converging
  toward a location at around $z=6$~Mm where the null point is expected; in
  fact, the field intensity at that point is seen to be a minimum compared to
  the surroundings. This is a standard fan-spine null configuration; the
  upper spine and neighboring field lines arch over the active region and
  link to the large negative polarity to the East of the image. These field
  lines provide a path for Jet 1 and Jet 2 towards the East.

 Figure~\ref{NFFF} shows a blow-up of the fan-spine structure. The
  region harboring the null point is marked with a green arrow and a star; the figure also indicates the approximate location of the inner spine (linking the null point to the parasitic
  negative polarities below it) and the outer spine that joins the
  null point to the large negative polarity to the East of the Active
  Region. The SST observations show that flux emergence is continually
  taking place in the region underneath the fan surface; moving magnetic patches are detected in the space between the opposite polarities N1-N2-N3 / P1-P2-P3 (see Section~\ref{sec:magnetic_flux_emergence} and the accompanying animation to Fig.~\ref{SST_HMI}). All of this is indicative of a reorganization of the magnetic structure in the low atmosphere underneath the fan surface that is likely to lead to the production of jets 1 and 2.

Understanding the origin of jet 3 is more complicated. The weak
  visibility in the HMI vector magnetic maps of the magnetic field polarities
  to the south of the region analyzed in the previous paragraph prevents the
  identification of a second null point (or region with large connectivity
  gradient) in that area.  On the other hand, the SST FOV contains that
  region and can resolve the negative polarities N4, N5, but it does not
  cover the full active region
(see~Fig.~\ref{FigHMI_SST}); so the SST data
do not provide a flux-balanced magnetic map on which an extrapolation could
be based. Yet, one can well imagine that, after the launching of jets 1 and
2, there is a drift of the reconnection site toward the south, which is also supported by the moving brightening observed in the hot AIA
94~\AA\ channel. Further, 
going back to Fig.~\ref{exp1}, the long yellow extrapolated
field lines were traced from a region to the south of (and near to)
the fan-spine configuration discussed
above; one sees that they first have 
a southward-pointing orientation that roughly 
follows the open loop path in AIA 171~\AA, and are co-spatial with
the path for the Jet 3. This could explain the abrupt change of
  orientation of Jet 3 compared with that of the two previous jets.

\section{Discussion and conclusions}
\label{sec:discuss}
 We present three recurrent solar jets observed in AR NOAA 13038 on June 24, 2022, using SDO and SST observations. The novelty of this work lies in the coordination of SDO observations with SST. 
 It is exceptional that we were able to resolve the fine-scale chromospheric structures in the jet environment using high spatial and temporal observations. 
The main observational findings of the paper are as follows:
\begin{enumerate}
    \item  
   Due to the continuous emergence of small parasitic magnetic fluxes at the edge of the main positive sunspot in the active region over the course of a day, the negative polarities are surrounded by positive polarities, as observed with CRISP/SST. This creates a magnetic configuration featuring a dome, a fan, and a null point.
    Due to the continuous emergence of new flux at the edge of the main positive sunspot in the active region over the course of a day, a collection of negative polarities appear next to the sunspot surrounded by positive flux patches, as observed with CRISP/SST. The presence of these parasitic polarities lead to a magnetic configuration featuring a dome, a fan and a null point.
    
    \item  
    The EUV images from AIA (171 \AA, 304 \AA) show a dark area surrounded by an oval of brightenings at this location. This bright oval corresponds to the base of the dome. The dark area might be interpreted as a mini filament; however, SST images in H$\alpha$ and H$\beta$ reveal an AFS instead of a filament. The filaments in this system are fine fibrils oriented from the positive polarities (P1, P2, P3) to the negative polarities (N1, N2, N3), similar to the positive threads seen in CRISP magnetic maps. They form loops or arcades with downflows at both ends, as illustrated in a sketch of emerging flux by \citet{Schmieder2014}. These fibrils are not perpendicular to the PIL but exhibit a shear, which introduces free energy.

    \item  
    The Doppler shifts obtained using H$\alpha$ (CRISP) and  H$\beta$ (CHROMIS) spectra  clearly show  that  a large flux rope, with width comparable to the width  of the dark area, is swirling over the AFS from North to South. This large twisted flux rope  forms because of  the shear during the  flux emergence phase.

    \item 
    The oval-shaped brightening observed in EUV lines corresponds to the flare ribbon seen in H$\alpha$ observations. Along this ribbon, several bright points 
    in EUV lines are observed, which could be due to cancelling flux. However, these points do not correspond to the bases of the jets.
    
    \item
    In H$\alpha$, following the swirling flux rope, another system of AFS developed from North to South, but with a different orientation. These could either be cooling post-flare loops or another AFS. However, if reconnection occurs at the null point, it is likely to observe post-flare loops.
    \end{enumerate}

Magnetic flux emergence is crucial for injecting magnetic energy into a system and for stressing the preexisting magnetic field through shearing. Magnetic reconnection can occur continuously and at different heights within the solar atmosphere. As a result of reconnection, flux cancellation can occur between the emerging parasitic polarities and the opposite-polarity preexisting region. In our current observational case, we conjecture that the jets occurred due to long-term continuous magnetic flux emergence, which induced shear between the pre-existing magnetic field and the newly emerging magnetic patches. In such scenarios, magnetic flux emergence might inject Poynting flux into the system and drive magnetic flux cancellation and ejection of jets. Initially, the continuous motion of negative and positive polarities created a sheared region, leading to the formation of a twisted flux rope.
Two jets were ejected to the East, and later, another jet (Jet3) was observed being ejected towards the South. Similarly, we also observed the drifting of a brightening at the jet base towards the South in the hot EUV channel, which could correspond to a moving reconnection site.

For this small and complex region, the NFFF extrapolation confirms the existence of a null point, corresponding to the jet flow towards the East. From the extrapolation, the height of the null point is around $z = 6$~Mm, which is consistent with the ranges given in the literature \citep{Galsgaard2017, Joshi2020}. In this present observation, it is not possible to establish accurately the existence of a second null point in the South, as the HMI magnetogram only shows weak polarities there, and the area is out of the SST FOV.

The change of direction of the jets from East to South could be a result of the displacement of the reconnection site to the southern region where the initial sections of the magnetic loops are oriented North-South, as seen in projection in the observation. 
From these EUV observations, we have estimated a possible drift of the reconnection site (brightening in the hot EUV channels) of 20 Mm in 18 min, implying an average velocity of 18.5 km s$^{-1}$. This is larger than the velocities inferred for null point displacements related to Coronal Bright Points from observations \citep{Galsgaard2017} and simulations \citep{Nobrega2022}. However, it is noteworthy that the magnetic field topology analyzed in the present paper is more complex.
The direction of the jets would depend on the site of the magnetic reconnection. The earlier reconnection events (Jet1 and Jet2) have potentially altered the position of the reconnection site and the orientation of the magnetic field lines passing through the reconnection region. Such a reconnection-driven shifting can easily happen if the spine (in the case of a null point) or the high-Q flux tube (in the case of a QSL) are located within a region where all field lines strongly diverge from the reconnection region. With such geometry, it is quite simple to shift the spine or the high-Q tube to nearly any position within the "fanning-out" diverging flux system. 

 Using the SST observations, we identified the presence of an AFS at the base of the jets, typically representing the chromospheric manifestations of emerging flux regions. This AFS appears as dark structures in EUV lines and can be identified as mini filaments in the absence of high-resolution observations. In this regard, we look forward to high-resolution observations  
from both existing and upcoming ground-based instruments, like DKIST \citep{DKIST2020} and EST \citep{EST2022}. Given the results of the extrapolation, and the changing direction of the jets, we conclude that, during the evolution of the jets, the reconnection site could have moved from one location to another. In general, during jet events or in confined flare events, the null point (reconnection site), may actually be able to move and eventually involve completely different magnetic loops in a new magnetic reconnection event \citep{Masson2009} or sometimes may show a persistent nature \citep{Cheng2023}. This can also be interpreted as `interchange reconnection' which gives rise to two reconnection products and sets a new footpoint connection for the coronal magnetic field lines \citep{Sterling2015}. In future high resolution ground based observations from the photosphere coordinated with space instruments for coronal observations \citep[for example MUSE,][]{MUSE2022,MUSE2022b} will provide a clearer picture about the moving reconnection site scenario for solar jets.

\begin{acknowledgements}
We thank to the referee for the comments  which help to improve the manuscript.
This research is supported by the Research Council of Norway, project number 325491, % ISSRESS (Luc)
 through its Centres of Excellence scheme, project number 262622, % RoCS (Reetika + Luc + Daniel)
 and the European Research Council through the Synergy Grant number 810218 (``The Whole Sun'', ERC-2018-SyG). This work benefited from discussions at the Instituto de Astrof\'isica de Canarias under the ``IAC early career visitor program'' and at the International Space Science Institute (ISSI) in Bern, through ISSI International Team project \#535 ``Unraveling surges: a joint perspective from numerical models, observations, and machine learning''.
%observations SST
The Swedish 1-m Solar Telescope is operated on the island of La
Palma by the Institute for Solar Physics of Stockholm University in the Spanish
Observatorio del Roque de Los Muchachos of the Instituto de Astrof\'isica
de Canarias. The Institute for Solar Physics is supported by a grant for research
infrastructures of national importance from the Swedish Research Council (registration
number 2017-00625). 
%SDO
SDO observations are courtesy of NASA/SDO and the AIA, EVE, and HMI science teams.
%ADS
We made use of NASA’s Astrophysics Data System Bibliographic Service.
\end{acknowledgements}

\bibliography{reference}
\bibliographystyle{aa}
\end{document}